\documentclass[aps,preprint]{revtex4}
\usepackage{amssymb}

\usepackage{amsmath}
\usepackage{graphicx}

\begin{document}
\title{Nuclear halo and its scaling laws}
\author{Z.H. Liu, X.Z. Zhang, H.Q. Zhang}
\affiliation{China Institute of Atomic Energy, Beijing 102413, China}

\begin{abstract}
We have proposed a procedure to extract the probability for valence particle
being out of the binding potential from the measured nuclear asymptotic
normalization coefficients. With this procedure, available data regarding
the nuclear halo candidates are systematically analyzed and a number of halo
nuclei are comfirmed. Based on these results we have got a much relaxed
condition for nuclear halo occurrence. Furthermore, we have presented the
scaling laws for the dimensionless quantity $<r^{2}>/R^{2}$ of nuclear halo
in terms of the analytical expressions of the expectation value for the
operator $r^{2}$ in a finite square-well potential.
\end{abstract}

\date[Date text]{May 29, 2003}
\maketitle

\medskip\ \thinspace \thinspace \thinspace \thinspace \thinspace \thinspace
\thinspace \thinspace \thinspace \thinspace \thinspace \thinspace \thinspace
\thinspace \thinspace \thinspace \thinspace \thinspace \thinspace \thinspace
\thinspace \thinspace \thinspace \thinspace \thinspace \thinspace \thinspace
\thinspace \thinspace \thinspace \thinspace \thinspace \thinspace \thinspace
\thinspace \thinspace \thinspace \thinspace \thinspace \thinspace \thinspace
\thinspace \thinspace \thinspace \thinspace \thinspace \thinspace \thinspace
\thinspace \thinspace \thinspace \thinspace \thinspace \thinspace \thinspace
\thinspace \thinspace \thinspace \thinspace \thinspace \thinspace \thinspace
\thinspace \thinspace \thinspace \thinspace \thinspace \thinspace \thinspace
\thinspace \thinspace \thinspace \thinspace \thinspace \thinspace \thinspace
\thinspace \thinspace \thinspace \thinspace \thinspace \thinspace 1.
INTRODUCTION

\medskip Nuclear halo is a threshold phenomenon\cite{refHa95}. As the
binding energy becomes small, the wave function of valence particle extends
more and more outward if the centrifugal barrier is small ($l\leq 1$).
Eventually, this leads to the wave function penetrating substantially beyond
the range of nuclear force as the binding energy approches zero, i.e.,
occurrence of nuclear halo.

The occurrence conditions for nuclear halo have already been discussed in
detail in the literature\cite{refRi92,refFe93,refFe94,refJe00}. Jensen and
Riisager\cite{refJe00}, recently, proposed the necessary conditions,

\begin{equation}
B_{p}<270\frac{A^{2}}{Z^{2}}\text{ keV \thinspace
\thinspace \thinspace for s-states,}  \label{eq-1}
\end{equation}

\begin{equation}
Z<0.44A^{4/3}\text{ \thinspace \thinspace \thinspace
\thinspace for p-states,}  \label{eq-2}
\end{equation}

\begin{equation}
B_{p}<2A^{-2/3}\text{ MeV \thinspace \thinspace
\thinspace for both s- and p-states,}  \label{eq-3}
\end{equation}%
where $B_{p}$ is the binding energy of valence particle. Eq.(\ref{eq-3})
overrules Eq.(\ref{eq-1}) and becomes decisive for all $A$ for s-states as
they pointed out. Most of halo nuclei experimentally observerd as well as
theoretically predicted are s-wave halo. Hence, Eq.(\ref{eq-3}) is the most
important conditions for halo occurrence.

Among the nuclear halo candidates, however, only the well-known s-wave halos
in $^{11}Li$, $^{11}Be$ and the p-wave halo in $^{11}Be$ with the $%
B_{p}A^{2/3}$ values of $1.48$ $MeV$, $2.49$ $MeV$ and $0.89$ $MeV$,
respectively, are below or close to the limit set by Eq.(\ref{eq-3}). As
will be seen , the others have binding energies much larger than this
requirement. Therefore, it might be necessary to make a systematic inquiry
concerning the occurrence conditions and scaling laws of nuclear halo. To
this end, we have first to select, if possible, a proper experimental
observable which can be employed to characterize the spatial extension of
valence particle density. Direct measurements of the nucleonic density
distribution should provide a visualized picture of nuclear skin and halo
stuctures, therefore will be very interesting. Over the years a large
variety of experimental methods have been developed, using leptonic probs
(as electrons, muons, etc.) for investigating nuclear charge distributions,
and hadronic probs (as protons, $\alpha -$particles, pions, etc.) for
exploring the distributions of nuclear matter \cite{refEg02}. All these
methods were applied successfully for many years for the study of stable
nuclei. Nowadays, study on the structures of exotic nuclei with many
possible methods has become a new and exciting field of research. Recently,
proton-nucleus elastic scattering at intermediate energies was applied for
obtaining accurate and detailed information on the size and radial shape of
halo nuclei \cite{refAl97}. Instead of density distribution, the probability
of a valence particle being out of the binding potential may be a suitable
quantity to assess the degree of the spatial extension. In the following, we
will present a procedure to extract the probability from experimental data
in a more reliable way, and try to get the occurrence conditions and scaling
laws of nuclear halo, which might be in line with experimental observations.
In this paper, we will limit ourselves to an investigation on two-body
nuclear halos.

\medskip \thinspace \thinspace \thinspace \thinspace \thinspace \thinspace
\thinspace \thinspace \thinspace \thinspace \thinspace \thinspace \thinspace
\thinspace \thinspace \thinspace \thinspace \thinspace \thinspace 2.
PROBABILITY OF VALENCE PARTICLE BEING OUT OF THE BINDING POTENTIAL

The probability for valence particle being out of the binding potential
radius $R$ can be evaluated by,

\begin{equation}
P(R,r_{0},a_{0})=\frac{\int_{R}^{\infty }r^{2}\phi _{nlj}^{2}(r)dr}{%
\int_{0}^{\infty }r^{2}\phi _{nlj}^{2}(r)dr}\text{,}  \label{eq-4}
\end{equation}%
where $\phi _{nlj\text{ }}$represents the normalized sigle-particle radial
wave function in the ($nlj$) bound state, $r_{0}$ and $a_{0}$ are the radius
and diffuseness parameters of the potential,and the binding potential radius
$R$ is a equivalent square-well potential radius which can be derived from
the measured mean-square core radius by $R^{2}=\frac{5}{3}\left(
<r^{2}>_{c}+4\right) $ $fm^{2}$ \cite{refFe93,refRi00}. We have explicitly
indicated the dependence of the probability $P$ on $R$, $r_{0}$ and $a_{0}$.
At the asymptotic distance $\phi _{nlj}$ behaves like,

\begin{equation}
\phi _{nlj}\simeq b_{lj}\frac{W_{-\eta ,l+1/2}(2kr)}{r},  \label{eq-5}
\end{equation}%
where $W_{-\eta ,\text{ }l+1/2}(r)$ is the Whittaker function, $k=\sqrt{2\mu
B_{p}/\hslash ^{2}}$ is the wave number, and $\mu $ and $\eta $ are the
reduced mass and the Coulomb parameter for system $\left( A_{c}+N\right) $.
In the neutron case where $\eta =0$, Whittaker function is related to the
modified Bessel function $K_{l+1/2}(kr)$ as $W_{0,l+1/2}(2kr)=\sqrt{kr/\pi }%
K $\thinspace $_{l+1/2}(kr)$\cite{refAb70}. In Eq. (\ref{eq-5}), $b_{lj}$ is
the single-particle asymptotic normalization coefficient ($ANC$) defining
the amplitude of the single-particle wave function in the asymptotic region.
The single-particle $ANC$ and the single-particle spectoscopic factor $%
S_{lj}^{(sp)}$ are related to the nuclear $ANC,C_{A_{c}Nlj}^{A},$ by\cite%
{refMu01},

\begin{equation}
C_{A_{c}Nlj}^{A}=b_{lj}(S_{lj}^{(sp)})^{1/2}\text{.}  \label{eq-6}
\end{equation}%
for virtual process $A\rightleftarrows A_{c}+N$. Here $A,$ $A_{c}$ and $N$
stand for the nucleus, its core nucleus and the valence particle,
respectively. Nuclear $ANC$ is an experimentally measurable and independent
of parameters of the potential \cite{refMu01,refLiu01}.

The probability calculated with Eq.(\ref{eq-4}) is only theoretical value
and potential-parameter dependent. Eq. (\ref{eq-6}) sets a restraint on the
single-particle wave function $\phi _{nlj}(r).$The amplitude of $\phi
_{nlj}(r)$ at the asymptotic distances, $b_{lj}$, is fixed as long as the
values of $C_{A_{c}Nlj}^{A}$ and $S_{lj}^{(sp)}$ are given. However, the
potential parameters $r_{0}$ and $a_{0},$ hence the single-particle wave
functions are not determined uniquely. We search the potential parameters in
such a way that the quantity,

\begin{equation}
\chi _{p}^{2}=\sum_{r_{i}=R_{N}}^{40fm}\left( S_{lj}^{1/2}\phi _{nlj}\left(
r_{i}\right) -C_{A_{c}Nlj}^{A}\frac{W_{-\eta ,l+1/2}\left( 2kr_{i}\right) }{%
r_{i}}\right) ^{2}  \label{eq-7}
\end{equation}%
becomes minimum. This is because $\phi _{nlj}(r)$ has the asymptotic
behaviour specified by Eq. (\ref{eq-5}) and Eq. (\ref{eq-6}) when $\chi
_{p}^{2}$ turns into minimum. We extract the probability from the
experimental data of $C_{A_{c}Nlj}^{A}$ and $S_{lj}^{(sp)}$ as follows.
First, single-particle wave function is calculated with the Woods-Saxon
potential. The depth of the potential is adjusted to reproduce the binding
energy $B_{p}$. The values of diffuseness parameter $a_{0}$ are choosen in
the range of $0.50-0.70$ fm. For each fixed $a_{0}$ , the radius parameter $%
r_{0}$ is varied in a small step till the minimum in $\chi _{p}^{2}$ is
reached. In the calculation of $\chi _{p}^{2}$, the summation runs from $%
r_{i}=6.0$ fm to 40 fm in step of 0.1 fm . We find that the value of $\chi
_{p}^{2}$ does not change as long as $R_{N}\geq 6$ fm. Then, for each pair
of $\left( r_{0},a_{0}\right) $ the probability for the valence particle
being out of the binding potential radius R is calculated with Eq. (\ref%
{eq-4}). Alternatively, one may take the radius, $R_{ws}=r_{0}A_{c}^{1/3%
\,}[1+5/3(\pi a_{0}/(r_{0}A_{c}^{1/3\,}))^{2}]^{1/2\,}$\cite{refFe94}
,\thinspace as binding potential radius instead of $R$. We find that the
value of $R$ and the values of $R_{ws}$ at the minima in $\chi _{p}^{2}$%
\thinspace \thinspace are nearly\thinspace the same. Hence, the
probabilities calculated with Eq. (\ref{eq-8}) below are nearly identical
whether $R$ or $R_{ws}$ is used. Besides, the results are the same within
the experimental error when $S_{lj}^{(sp)}$ changes 10\%. Because the
relative uncertainty of the experimental $S_{lj}^{(sp)}$ is usually less
than 10\%, therefore, the results below are reliable within the accuracy of
\thinspace $S_{lj}^{(sp)}$. Figure 1 shows the $\chi _{p}^{2}$ as a function
of the probability $P$ for the $2s$ excited state in $^{13}C$. In this
calculation, $C_{A_{c}Nlj}^{A}=1.84\pm 0.16fm^{-1/2}$ \cite{refLiu01} and
\thinspace $S_{lj}^{(sp)}=0.95$ \cite{refJa83} are used. It can be seen from
the figure that the\thinspace minimum in $\chi _{p}^{2}$ is very deep with a
very small dispersion in the probability $P$. This is actually due to the
fact that $\chi _{p}^{2}$ reaches its minimum value when Eq. (\ref{eq-6}) is
satisfied. Therefore, in terms of the experimentally measured nuclear
asymptotic normalization coefficients $C_{A_{c}Nlj}^{A}$, we can calculate
the average value of the probability $P$ by,

\begin{equation}
P=\sum w_{i}P(R,r_{0i},a_{0i})\text{,}  \label{eq-8}
\end{equation}%
with

\begin{equation}
w_{i}=\frac{(\chi _{p}^{2}\left( r_{0i},a_{0i}\right) )^{-1}}{\sum \left(
\chi _{p}^{2}\left( r_{0i},a_{0i}\right) \right) ^{-1}}\text{.}  \label{eq-9}
\end{equation}%
The summation runs over the minimum points in $\chi _{p}^{2}$ for different $%
a_{0}$. It is worth to note that the probability obtained in this way is
nearly parameter independent. This is an interesting and meaningful result
within the reach of present experimental knowledge.

By means of the above procedure, the probability for the valence particle
being out of the binding potential have been calculated for a number of
nuclei. The values listed in Table 1 are the weighted average probability $P$
extracted from different experimental sources. The results are plotted in
Figs. 2(a) and (b) as a function of $2\mu B_{p}R^{2}/\hbar ^{2}$ and $%
B_{p}A^{2/3}$, respectively. Riisager et al.\cite{refRi94,refHa95,refJe00}
suggest a criterion for quantitative assessment of halos, i.e., the valence
particle has large probability, say > 50\%, of being out of the
nuclear binding potential radius. According to this criterion, the nuclei
under considerations are all s-wave halo in the ground state (solid points)
or in the excited states (open circles). We find from Fig. 2(b) that halo
may be able to occure for

\begin{equation}
B_{p}A^{2/3}<10\text{ }MeV  \label{eq-10}
\end{equation}%
which is much relexed than the one given by Eq. (\ref{eq-3}).

We have calculated the probability for the valence neutron in 2s-state using
Woods-Saxon potential with normal parameters $r_{0}=1.27$ $fm,$ \thinspace $%
a_{0}=0.67$ $fm$, $U_{0}=\left( 50-32\frac{N-Z}{A}\right) $ $MeV$ for the
nuclei with $N=Z$ and $(N-Z)/A=1/3$, respectively. The binding potential
radius $R$ used here is also calculated by $R^{2}=\frac{5}{3}\left(
<r^{2}>_{c}+4\right) $ $fm^{2}$ and $<r^{2}>_{c}=(r_{0c}A^{1/3})^{2}$ ,the $%
r_{0c}$ is obtained by fitting to the experimental $rms$ radii of light
nuclei.The calculated results are also plotted in Fig. 2 as the solid and
dashed lines. Basically, they are in agreement with the experimental data.
In order to examine the effects of larger diffuseness, we have plotted in
Fig. 2(a) the probabilities for the Woods-Saxon potential with the same
parameters as those used above except for doubling the diffuseness, $%
a_{0}=1.34$ $fm$. As shown by the dash-dotted and dotted lines, these
calculations overpredict the experimental data. It means that using a very
large diffuseness in potential may not correspond to the realistic situation
for nuclei with weakly bound neutrons.

\medskip The probability of a valence particle being out of the square-well
potential is,

\begin{equation}
P=\frac{1}{\chi +1}\left( 1-\frac{\chi ^{2}}{\xi _{0}^{2}}\right) \text{ \ \
}l=0,  \label{eq-11}
\end{equation}

\begin{equation}
P=\frac{\chi +2}{\chi ^{2}+3\chi +3}\left( 1-\frac{\chi ^{2}}{\xi _{0}^{2}}%
\right) \text{ \ \ }l=1.  \label{eq-12}
\end{equation}%
For $l$=$2$,

\medskip\ $\,\,\,\,\,\,\,\,\,\,\,\,\,\,\,\,\,\,\,\,\,\,\,\,\,\,\,\,\,\,\,\,%
\,\,\,\,\,\,\,\,\,\,\,\,\,\,\,\,\,\,\,\,\,\,$%
\begin{equation}
P=\frac{\chi ^{3}+6\chi ^{2}+12\chi +6}{(\chi +1)(\chi ^{3}+6\chi
^{2}+15\chi +15)}(1-\frac{\chi ^{2}}{\xi _{0}^{2}})\rightarrow 0.4\text{
\thinspace \thinspace \thinspace \thinspace }(\chi \rightarrow 0),
\label{eq-13}
\end{equation}%
where

\begin{equation}
\chi =R\sqrt{\frac{2\mu B_{p}}{\hbar ^{2}}},  \label{eq-14}
\end{equation}%
and

\begin{equation}
\xi _{0}=R\sqrt{\frac{2\mu U_{sq}}{\hbar ^{2}}}.  \label{eq-15}
\end{equation}%
In the above equations, $U_{sq}$ is the depth of the square-well potential.
The dash-double-dotted line in Fig. 2(a) illustrats the square-well
potential predictions for 2s-state. We see that it underpredicts the
experimentally extracted data, implying the important roles of the potential
diffuseness on the probability being out of the binding potential radius.

\medskip

\medskip\ \thinspace \thinspace \thinspace \thinspace \thinspace \thinspace
\thinspace \thinspace \thinspace \thinspace \thinspace \thinspace \thinspace
\thinspace \thinspace \thinspace \thinspace \thinspace \thinspace \thinspace
\thinspace \thinspace \thinspace \thinspace \thinspace \thinspace \thinspace
\thinspace \thinspace \thinspace \thinspace \thinspace \thinspace \thinspace
\thinspace \thinspace \thinspace \thinspace \thinspace \thinspace \thinspace
\thinspace \thinspace \thinspace \thinspace 3. SCALING LAWS OF TWO-BODY
NUCLEAR HALO

In terms of nuclear $ANC,$ we can extract the root-mean-square $\left(
rms\right) $ radius of the probability distribution for valence particle in
the orbit $\left( nlj\right) $. It can be written as the contributions from
the interior and asymptotic regions \cite{refLiu01,refCa01},

\begin{equation}
<r^{2}>^{1/2}=\left[ S_{lj}\int_{0}^{R}r^{4}\phi _{nlj}^{2}(r)dr+\left(
C_{A_{c}Nlj}^{A}\right) ^{2}\int_{R}^{\infty }r^{2}W_{-\eta ,\text{ }%
l+1/2}^{2}\left( 2kr\right) dr\right] ^{1/2}.  \label{eq-16}
\end{equation}%
The first term in the equation is somehow parameter dependent, while the
second term is not. Moreover, in the case of weakly bound nuclei, the second
term gives more than 90\% contribution to the value of the $rms$ radius.
Thus the error introduced by the parameters is small in the cases under
consideration. The $rms$ radii of the valence particle have been calculated
in this way for the nuclei $^{11}$Be \cite{refOz01,refAj90,refFo99,refAu00},
$^{12}$B \cite{refLiu01}, $^{13,14,15}$C \cite%
{refLiu01,refOz01,refLiu02,refAj91}, $^{19}C$ \cite{refOz01,refBa00}. They
are listed in Table \ref{table} along with other parameters. Based on the
assumption of a core plus a valence neutron structure, recently, Ozawa
\textit{et} \textit{al. }\cite{refOz01} applied a Glauber-model analysis for
a few body system $\left( GMFB\right) $, and deduced spectroscopic factors $%
S_{lj}$ for some selected nuclei from the measured interaction cross
sections $\sigma _{I}$. With their parameters of the binding potential, we
calculate the single-particle wave function and obtain single-particle $ANC$
$b_{lj}$ in asymptotic region. Then, the nuclear $ANC$ can be obtained from
the deduced $S$-factor and the single-particle $ANC$ $b_{lj}$ with Eq. (\ref%
{eq-6}). The $rms$ radii for $^{11}$Be$,^{15,19}$C are evaluated by means of
the $GMFB$ analysis, and the results are also listed in Table \ref{table}.
In order to check the above results, the\ $rms$ radii of the valence
particle wave functions in these three nuclei are extracted by subtracting
the core contribution from the mean-square matter radius \cite{refRi00}:

\begin{equation}
<r^{2}>=\frac{A^{2}}{A_{c}A_{h}}<r^{2}>_{m}-\frac{A}{A_{h}}<r^{2}>_{c},
\label{eq-17}
\end{equation}%
where $A$, $A_{c}$ and $A_{h}$ are the total, core and valence particle mass
numbers of the system, respectively. These $rms$ radii are compared with the
other data in Table \ref{table}. Except for $^{15}$C, the $rms$ radii of
halo obtained with these three methods are in agreement within the
experimental errors.

\medskip Hamamoto and Zhang\cite{refHa98} have deduced the expressions for
the expectation value of the operator $r^{2}$ in a finite square-well
potential. The terms with $\xi _{0}^{4\text{ }}$in denominator in thier
expressions are negligible in magnitude as compared to the other terms for
the case of $\chi ^{2}<2$ which we are interested in. After omitting them,
we get the following scaling laws,

\begin{equation}
\frac{<r^{2}>}{R^{2}}=\frac{1}{\chi +1}\left[ \left( 1-\frac{\chi ^{2}}{\xi
_{0}^{2}}\right) (1+\frac{1}{\chi }+\frac{1}{2\chi ^{2}})+\chi (\frac{1}{3}+%
\frac{1}{2\xi _{0}^{2}}+\frac{\chi }{\xi _{0}^{2}})\right] \text{ \ \ }l=0,
\label{eq-18}
\end{equation}

\begin{equation}
\frac{<r^{2}>}{R^{2}}=\frac{1}{\chi ^{2}+3\chi +3}\left[ \left( 1-\frac{\chi
^{2}}{\xi _{0}^{2}}\right) \left( \frac{(\chi +1)^{2}}{3}+\chi +3+\frac{5}{%
2\chi }\right) +\frac{(\chi +1)^{2}}{2\xi _{0}^{2}}\right] +\frac{\left(
\chi ^{2}+2\chi +2\right) }{3\xi _{0}^{2}}\text{ \ \ }l=1  \label{eq-19}
\end{equation}%
Keeping the largest term of the above eqations in the limit $\chi
\rightarrow 0$, we will arrive at the scaling laws in Ref.\cite{refFe93}. It
should be kept in mind that the above laws depend on the quantum number $n$
through $\xi _{0}$. Here $n$ is the node number of the radial wave function
of valence particle. If halo is defined in terms of the requirement that the
experimental value of probability $P$ is greater than 50\%, or
approximatelly $\chi ^{2}\leq 1.8$ (see Fig.2(a)) ,we get the following
conditions from Eqs. (\ref{eq-18}) and (\ref{eq-19}) for nuclear halo
occurrence,

\begin{equation}
\frac{<r^{2}>}{R^{2}}\geq 1.5\text{ \ \ for }2s\text{ states,}  \label{eq-20}
\end{equation}

\begin{equation}
\frac{<r^{2}>}{R^{2}}\geq 1.9\text{ \ \ for }1p\text{ states.}  \label{eq-21}
\end{equation}%
Since the probability $P$ is less than 40\% for $l=2$, we come to the same
conclusion as Riisager et al \cite{refRi92} that halo is unlikely to occur
for the particle in the $d$ states.

In Fig. \ref{fig3}, the experimental data of $<r^{2}>/R^{2}$ are compared
with our scaling law as well as the predictions of the single-particle model
for the valence particle in $2s$ state. In order to have a better
statistics, we adopt the weighted average value of $<r^{2}>$ instead of the
individual experimental results. For the data without error, we assigned
them to 10\% of uncertainty for evaluating the weighted average.We see from
the figure that the scaling law Eq. (\ref{eq-18}) can account for the
available experimental data of halo candidates,though it is derived in a
finite square-well potential. Several authors \cite%
{refRi00,refFe93,refHa87,refJe01} have put forward their scaling laws. Being
of their $l$ and/or $n$ independent, we \thinspace do not present them in
the figure.

\medskip

\medskip\ \thinspace \thinspace \thinspace \thinspace \thinspace \thinspace
\thinspace \thinspace \thinspace \thinspace \thinspace \thinspace \thinspace
\thinspace \thinspace \thinspace \thinspace \thinspace \thinspace \thinspace
\thinspace \thinspace \thinspace \thinspace \thinspace \thinspace \thinspace
\thinspace \thinspace \thinspace \thinspace \thinspace \thinspace \thinspace
\thinspace \thinspace \thinspace \thinspace \thinspace \thinspace \thinspace
\thinspace \thinspace \thinspace \thinspace \thinspace \thinspace \thinspace
\thinspace \thinspace \thinspace \thinspace \thinspace \thinspace \thinspace
\thinspace \thinspace \thinspace \thinspace \thinspace \thinspace \thinspace
\thinspace \thinspace \thinspace \thinspace \thinspace \thinspace \thinspace
\thinspace \thinspace \thinspace \thinspace \thinspace \thinspace \thinspace
\thinspace \thinspace \thinspace \thinspace \thinspace \thinspace \thinspace
\thinspace \thinspace 4. SUMARRY

In summary, we have proposed a procedure to extract the probability for
valence particle being out of the binding potential from the measured
nuclear asymptotic normalization coefficients. With this procedure,
available data regarding the nuclear halo candidates are systematically
analyzed and a number of halo nuclei are comfirmed. Based on these results
we have got a much relaxed condition for nuclear halo formation as compared
to Ref.\cite{refJe00}. The effect of potential deffuseness on the
probability being out of the nuclear binding potential radius is also
discussed. In terms of the analytical expressions of the expectation value
for the operator $r^{2}$ in a finite square-well potential, we have
presented the scaling laws for the dimensionless quantity $<r^{2}>/R^{2}$ of
nuclear halo,which can account for the available experimental data of halo
candidates.\thinspace

\begin{acknowledgments}
This work was supported by the National Natural Science Foundation of China
under Grants No.10075077, 10105016, 10275092 and the Major State Basic
Research Development Programme under Grant No. G200007400.
\end{acknowledgments}

\medskip\ \thinspace \thinspace \thinspace
%TCIMACRO{\TeXButton{B}{\begin{table}[tbp] \centering}}%
%BeginExpansion
\begin{table}[tbp] \centering%
%EndExpansion
\caption{Deduced nuclear $ANC$, $probability$, $rms$ radii and $<r^{2}>/R^{2}$ for the
nuclear halo candidates.\label{table}}%
\begin{tabular}[t]{ccccccccc}
\hline
Nucleus & $J^{\pi }$ & $B_{p}$ & $C_{ANlj}^{B}$ & $probability$ & $%
<r^{2}>^{1/2}$ & $<r^{2}>_{av}^{1/2}$ & $<r^{2}>_{av}/R^{2}$ & Ref. \\
&  & $\left( keV\right) $ & $\left( fm^{-1/2}\right) $ & $(\%)$ & $\left(
fm\right) $ & $\left( fm\right) $ &  &  \\ \hline
$^{11}Be$ & $1/2^{+}$ & 504 & 0.81$\pm $0.05 & - & 6.68$\pm $0.43$^{a}$ & -
& - & \cite{refOz01} \\
&  &  & - & - & 6.65$\pm $0.31$^{b}$ & - & - & \cite{refOz01} \\
&  &  & 0.76$\pm $0.03 & - & 6.23$\pm $0.25 & - & - & \cite{refAj90} \\
&  &  & 0.78 & - & 6.40 & - & - & \cite{refAj90} \\
&  &  & 0.81 & - & 6.68 & - & - & \cite{refFo99} \\
&  &  & 0.78 & - & 6.44 & - & - & \cite{refAu00} \\
&  &  & - & 78.6$\pm $7.5 & - & 6.46$\pm $0.16 & 2.60$\pm $0.13 &  \\
$^{12}B$ & $1^{-}$ & 749 & 0.94$\pm $0.08 & 68.5$\pm $11.7 & 5.64$\pm $0.90
& 5.64$\pm $0.90 & 1.86$\pm $0.60 & \cite{refLiu01} \\
& $2^{-}$ & 1696 & 1.34$\pm $0.12 & 54.2$\pm $4.0 & 4.01$\pm $0.61 & 4.01$%
\pm $0.61 & 1.10$\pm $0.31 & \cite{refLiu01} \\
$^{13}C$ & $1/2^{+}$ & 1857 & 1.84$\pm $0.16 & 55.5$\pm $9.6 & 5.04$\pm $0.75
& 5.04$\pm $0.75 & 1.55$\pm $0.47 & \cite{refLiu01} \\
$^{14}C$ & $0^{-}$ & 1274 & 1.54$\pm $0.09 & 64.0$\pm $7.5 & 5.78$\pm $0.36
& 5.78$\pm $0.36 & 1.97$\pm $0.26 & \cite{refLiu02} \\
& $1^{-}$ & 2083 & 1.84$\pm $0.11 & 56.1$\pm $6.7 & 4.57$\pm $0.30 & 4.57$%
\pm $0.30 & 1.34$\pm $0.18 & \cite{refLiu02} \\
$^{15}C$ & $1/2^{+}$ & 1218 & 1.05$\pm $0.22 & - & 3.65$\pm $0.82$^{a}$ & -
& - & \cite{refOz01} \\
&  &  & - & - & 4.59$\pm $1.02$^{b}$ & - & - & \cite{refOz01} \\
&  &  & 1.40 & - & 5.40 & - & - & \cite{refAj91} \\
&  &  & 1.49$\pm $0.15 & - & 5.86$\pm $0.60 & - & - & \cite{refLiu02} \\
&  &  & - & 60.4$\pm $6.6 & - & 5.15$\pm $0.34 & 1.55$\pm $0.21 &  \\
$^{19}C$ & - & 240 & 0.57$\pm $0.19 & - & 7.87$\pm $1.49$^{a}$ & - & - & %
\cite{refOz01} \\
&  &  & - & - & 7.63$\pm $2.46$^{b}$ & - & - & \cite{refOz01} \\
&  &  & 0.55$\pm $0.07 & - & 7.07$\pm $0.50 & - & - & \cite{refBa00} \\
&  &  & - & 80.8$\pm $19.9 & - & 7.17$\pm $0.47 & 2.58$\pm $0.34 &  \\ \hline
\end{tabular}%
\medskip \medskip \medskip
%TCIMACRO{
%\TeXButton{notation}{$^{a}$ {\footnotesize Deduced with GMFB method. \thinspace \thinspace }$^{b}$
%{\footnotesize Calculated with Eq. \ref{eq-17}}.} }%
%BeginExpansion
$^{a}$ {\footnotesize Deduced with GMFB method. \thinspace \thinspace }$^{b}$
{\footnotesize Calculated with Eq. \ref{eq-17}}.
%EndExpansion
\thinspace \thinspace \thinspace \thinspace \thinspace \thinspace \thinspace
\thinspace \thinspace \thinspace
%TCIMACRO{\TeXButton{E}{\end{table}}}%
%BeginExpansion
\end{table}%
%EndExpansion
\clearpage
\begin{figure}[h]
%TCIMACRO{
%\TeXButton{includegraphics}{\includegraphics[scale=1.0,angle=-90.]{Fig1.eps}} }%
%BeginExpansion
\includegraphics[scale=0.5,angle=-90.]{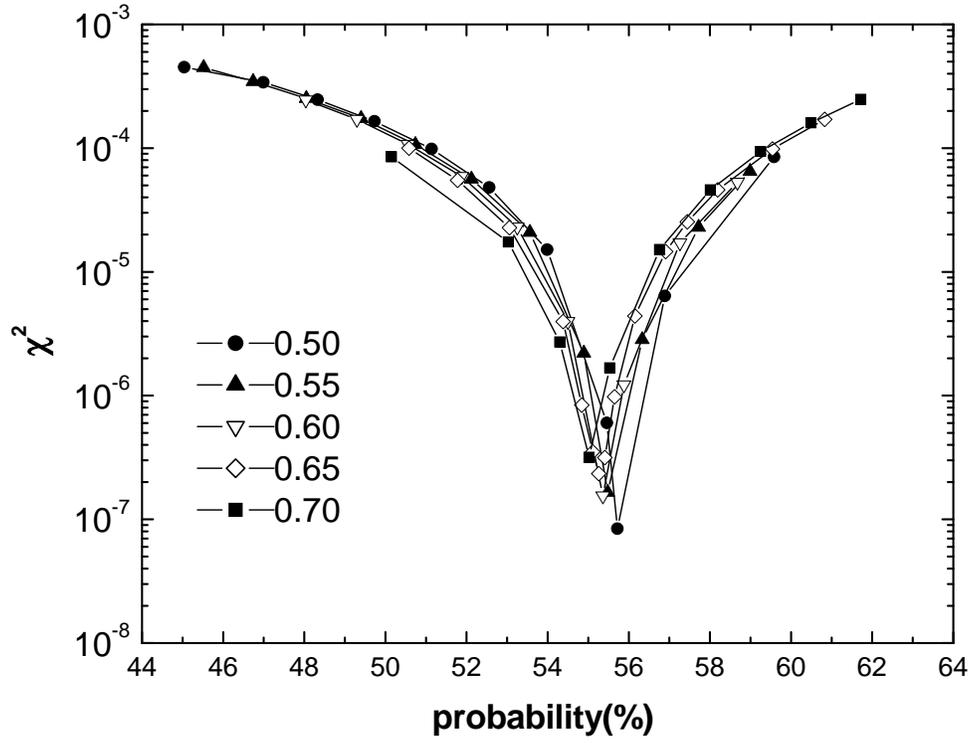}
%EndExpansion
\vspace*{0.5cm}
\caption{Dependence of the $\protect\chi _{p}^{2}$ on the probability for
valence particle being out of the binding potential $P$ for the $2s$ excited
state in $^{13}C$. Symbols are connected by a line for each $a_{0}$ value to
guide the eye.}
\label{fig1}
\end{figure}

\begin{center}
\medskip

\begin{figure}[h]
\includegraphics[scale=1.0,angle=-90.]{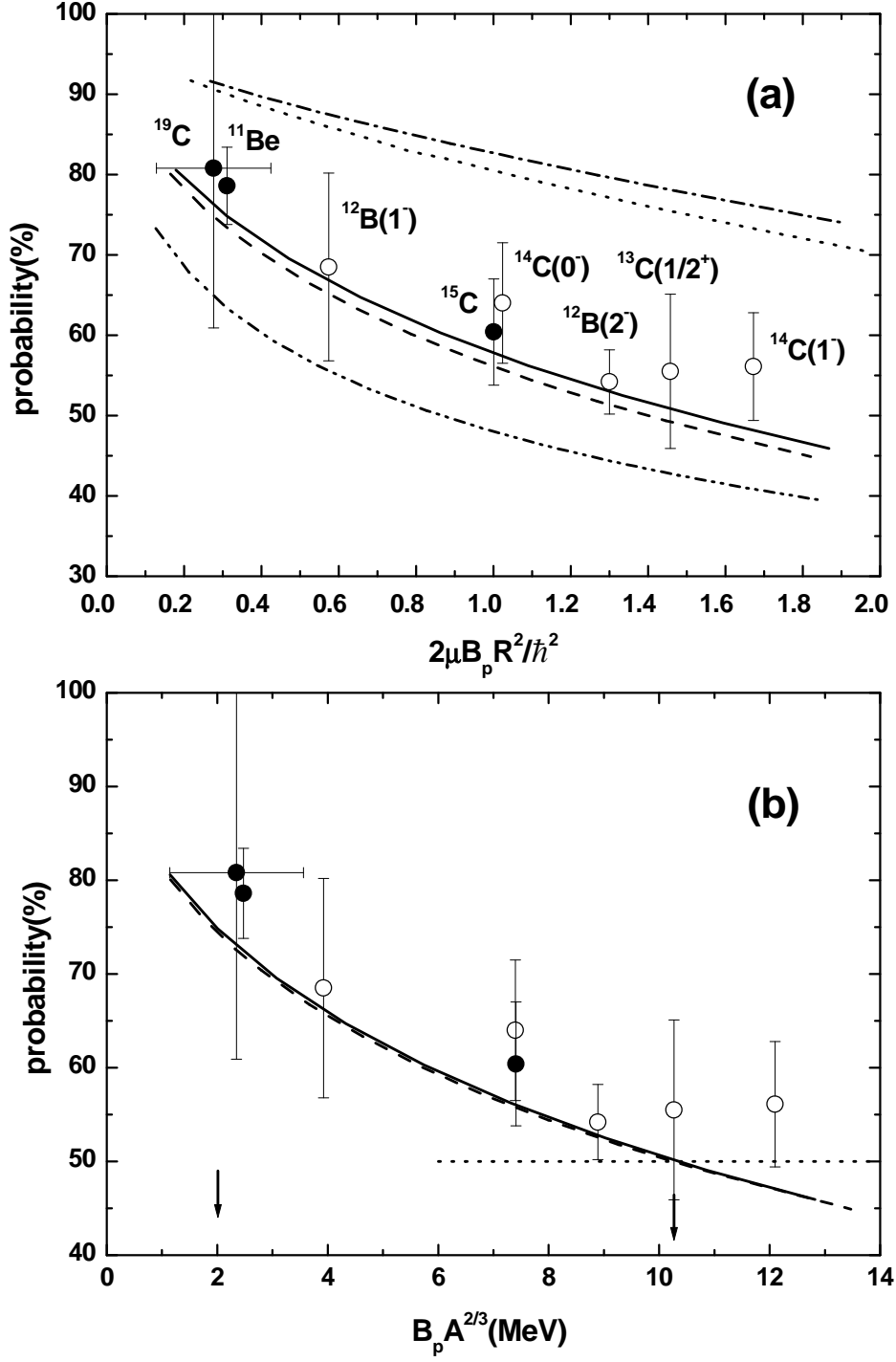} \vspace*{0.5cm}
\caption{Probability for valence particle being out of the binding potential
as a function of $2\protect\mu B_{p}R^{2}/\hbar ^{2}$ (a) and $B_{p}A^{2/3}$
(b). The solid points and open circles represent the $s$-wave halos in the
ground state and in the excited states, respectively. The lines show the
predictions of the single-particle models with Woods-Saxon potentials and
the square-well potential. The arrows in the panel (b) illustrate the
up-limits of $B_{p}A^{2/3}$ value set by Eq. (\ref{eq-3}), and Eq. (\ref%
{eq-10}). }
\label{fig2}
\end{figure}

\medskip

\begin{figure}[h]
\includegraphics[scale=0.5,angle=-90.]{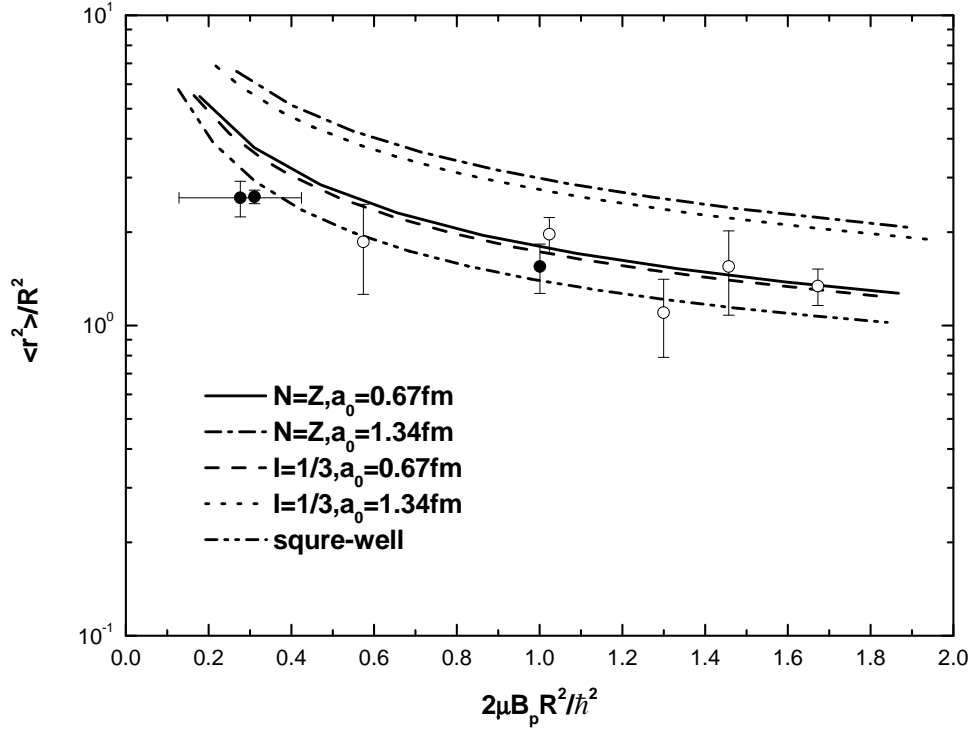} \vspace*{0.5cm}
\caption{Experimental data of $<r^{2}>/R^{2}$ vs $2\protect\mu %
B_{p}R^{2}/\hbar ^{2}$ for the valence particle in the $2s$ state. The
dash-double-dotted line is the scaling law of Eq. (\ref{eq-18}). The other
lines show the results of the single-particle model calculations with
Woods-Saxon potentials for the nuclei with N=Z and I=(N-Z)/A=1/3,
respectively. }
\label{fig3}
\end{figure}
\end{center}

\end{document}